# Informatics Modeling of High Tg Polymers: Assessing the Role of Processing versus Chemistry

Qinrui Liu [1], Scott R. Broderick [1,*]

[1]Department of Materials Design and Innovation, University at Buffalo, Buffalo, NY 14260, USA
* Correspondence: scottbro@buffalo.edu

**Abstract**

Despite the advances in structure-based modeling of polymer properties, accurately predicting glass transition temperature (Tg) is still challenging for polymers whose behavior is strongly influenced by intermolecular interactions and processing conditions. We previously developed a machine-learning model based on polymer topological descriptors to predict Tg. The model performed well and was based solely on the chemistry and structure of the polymer without any inclusion of processing parameters. In this work, we have extended that work by first applying that same model to a larger range of polymers and second by integrating processing parameters into the feature set. The chemistry-based model still demonstrates consistent predictive performance for most polymers, indicating that Tg is indeed primarily chemistry and structure driven and not strongly impacted by processing. However, several polymers exhibited deviations between predicted and experimental Tg values. Detailed analysis reveals that these differences are related to strong intermolecular interactions and processing-dependent factors, particularly for polymers prepared by solution casting and high temperature annealing. These results demonstrate that molecular topology provides a strong foundation for Tg prediction; however, this approach also screens out those classes of polymers for with processing conditions play an important role.

## I. Introduction

Polymers with enhanced thermal stability are highly desirable for applications in harsh and demanding environments. To accelerate the discovery and design of promising polymer candidates with targeted thermal and mechanical performance, as well as improved sustainability, we have conducted a series of studies applying machine-learning tools to polymer datasets. In our previous work [1,2], polymer topological descriptors were leveraged to construct predictive models, capturing an effective representation of the diverse chemical features of polymer monomers. Building on this framework, the present study extends our previous Tg-focused model to report additional observations related to polymer thermal behavior. The trained model was further applied to an added set of polymers with experimentally reported Tg values and corresponding processing conditions, providing further comparison between model predictions and experimental measurements. The difference between a chemistry/structure only model and a model with processing included provides quantitative information on the role of processing.

Polymer processing conditions are commonly regarded as having a limited impact on the glass transition temperature (Tg), which is generally treated as an intrinsic material property primarily governed by polymer chemistry [3–10]. Numerous studies have suggested that chain stiffness, backbone rigidity, and segmental mobility play an important role in determining Tg. From this perspective, Tg can be effectively correlated with the chemical building blocks of polymers, and meaningful structure–property relationships can be established using molecular descriptors or chemical fingerprints derived from polymer structures [6,9,10]. Other work has discussed the dependence of Tg on chemical factors such as molecular weight, specific side groups, copolymer composition, conjugation effects, and crystallinity [3–5]. In particular, several studies have demonstrated that Tg can be modeled and predicted directly from monomer level SMILES representations, which encode essential chemical information of polymers and serve as an effective input for data-driven models [8]. These findings all support the view that polymer chemistry plays the dominant role in governing Tg, while processing effects are often considered secondary in comparison [11].

However, there are notable exceptions in which specific processing conditions are associated with measurable variations in Tg.  These exceptions will be demonstrated in this paper and we show that for certain polymer systems, processing effects may play a more important role than is conventionally assumed. For example, polymers such as polybenzimidazole (PBI) and poly (phenylene oxide) are often processed via solution casting followed by high-temperature annealing, which can potentially lead to elevated Tg values [12-16]. Although solution casting itself does not intrinsically increase Tg, it can result in different polymer morphologies and residual solvent states that influence thermal behavior [17,18]. The final Tg is therefore dependent on multiple factors, including solvent choice, drying conditions, and polymer chemistry, and an increased Tg may be achieved by appropriate processing design.

Solution casting is a simple and low-cost technique that is widely used for laboratory-scale fabrication and large-area film preparation; however, it also has limitations, such as residual solvent retention and long drying times [18-21]. Notably, polymers commonly reported to be processed via solution casting rather than melt processing often share similar structural features, including heteroaromatic rings, rigid backbones, and strong hydrogen-bonding sites. Representative examples include PBI, Kevlar, PBO/PBT, and certain ladder-type polymers, which are generally considered more suitable for solution-based processing due to their high thermal stability and limited melt processability [22-27].  This work broadens the current understanding of structure–property relationships governing polymer Tg and highlights the potential importance of screening and testing processing-related factors in data-driven polymer design.

## II. Methodology

The predictive models used in this study were established in our previous work and are based on a tuned gradient boosting regression approach using ten selected topological descriptors. Descriptor calculation, model training, and hyperparameter optimization followed the workflow in the prior study. The trained model was applied to properties of 43 polymers extracted from an available handbook [28]. These polymers were selected because they have significant experimentally reported processing information is available for them from experimental resources, including processing temperature, processing time, processing pressure, and processing method category. This dataset was curated during our effort to further explore the potential effects of polymer processing conditions on Tg. Importantly, these polymers were not included in the training dataset used to develop the prior Tg model, which makes this additional evaluation particularly meaningful as an independent validation of the established model. The overall logic of this work is shown in Figure 1, highlighting the difference between the earlier work [1,2] and the extended dataset.

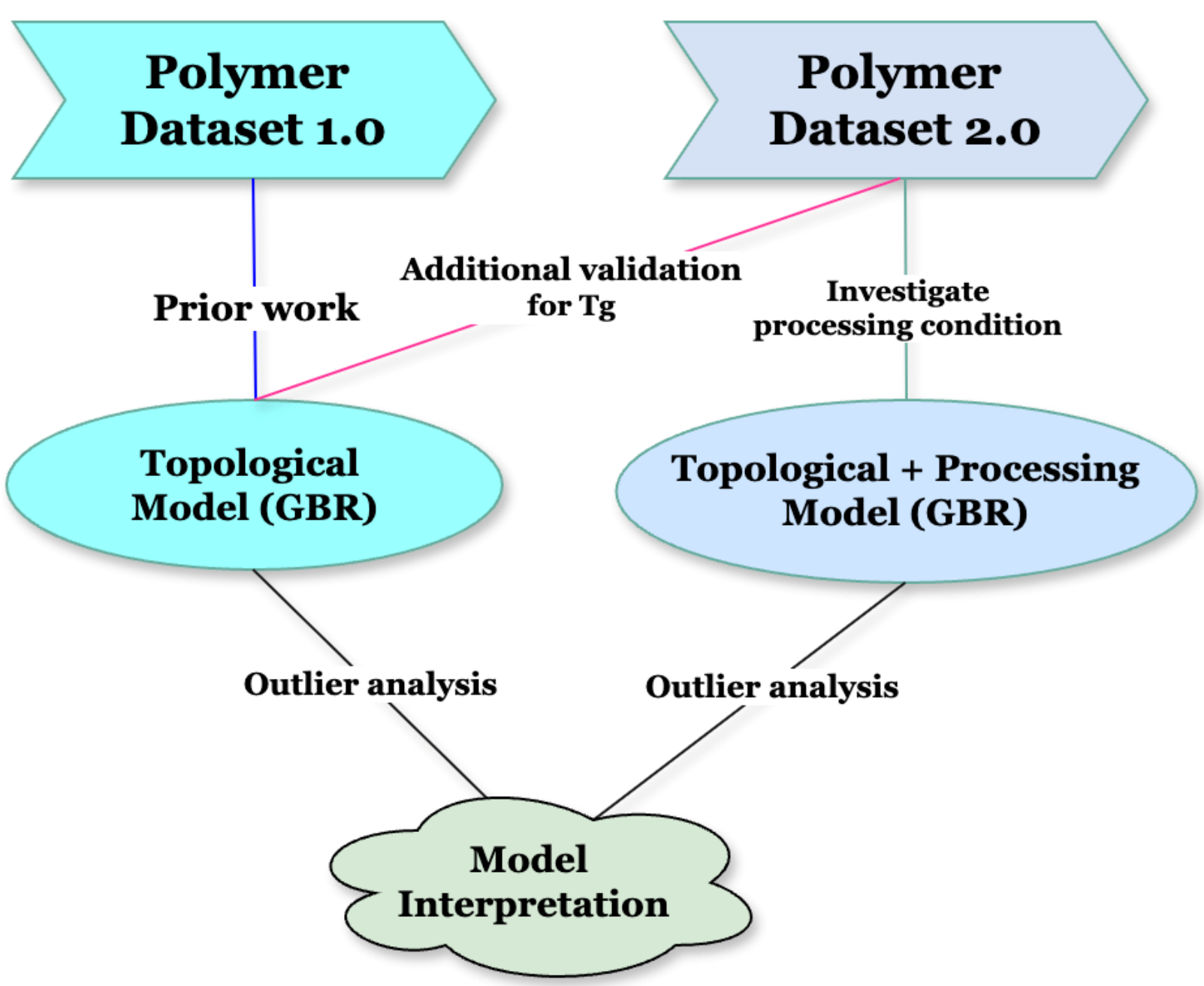

**Figure 1.** The logic overview of this study, showing the connection with the prior study and the key steps included. In the flowchart, the polymer dataset used in the prior work is represented by Polymer Dataset 1.0 and the additional polymers with as much as processing conditions available is represented by Polymer Dataset 2.0.

The Polymer Dataset 2.0 has 43 polymers included, with 35 temperature values, 15 pressure values, and 13 time values. The missing processing conditions were filled in through mean imputation. Given the small and sparse data, the processing information introduces significant uncertainty, but still provides some new information whose impact on models can be tracked.

The initial Tg model applied in this work was developed in our previous study and is based on a gradient boosting regression framework with explicit feature engineering. The model incorporates a selected set of ten topological descriptors, including 0X, N_rot, N_CH2, 0Xv, M, BB_index2, N_ether, N_aromaticRing, N_H, and BB_index1, which were designed to represent the diverse chemical characteristics of polymer repeat units. These features are defined in Table 1, with the features down selected from a larger initial feature space. Multiple regression models and feature combinations were initially screened during model development, and the gradient boosting model demonstrated the best overall performance and robustness. The final model achieved a training $R^2$ of 0.899 and a test $R^2$ of 0.869, with a test RMSE of 33.14 °C, indicating good predictive accuracy across a broad Tg range.

**Table 1.** The features used in the building of a model for Tg in our prior work, with the feature set developed and calculated for the relevant polymers.

| Polymer Features | How They Were Defined or Calculated |
|---|---|
| N_H | number of hydrogen atoms in one polymeric repeat unit |
| N_aromaticring | number of aromatic rings in one polymeric repeat unit |
| N_CH2 | number of $-CH_2$ in one polymeric repeat unit |
| N_ether | number of -O- in a polymeric repeat unit |
| M | mole weight of one polymer repeat unit(g/mol) |
| N_rot | total number of rotational degrees of freedom parameter (N_rot = the backbone rotational degrees plus the side group rotational degrees) |
| $^0\chi$ | the zeroth-order (atomic) connectivity indices (the first atomic index) |
| $^0\chi^V$ | the zeroth-order (atomic) connectivity indices (the second atomic index) |
| BB_index1 | backbone index1 is a steric hindrance parameter that reflects the flexibility of the polymer backbone structure, similar to the stiffness of the backbone. |
| BB_index2 | backbone index2 is a steric hindrance parameter that differentiates between backbone atoms with the same (δ/δV) values but different δ values, reflecting variations in the number of non-hydrogen neighbors around each backbone atom. |

## III. Results

In the prior work, an interpretable tree-based model was built based on a set of polymer topological features and provided a well-balanced performance for a limited experimental Tg dataset. However, in the model performance evaluation, we also acknowledged that some aspects related to intermolecular forces may not be fully captured by the feature set. In particular, four polymers which are vinyl acetal, vinyl butyral, p-phenylene terephthalamide, and sodium methacrylate were identified as underpredicted. This behavior was attributed to factors such as ring structures that impose conformational rigidity and reduce local rotational freedom, strong interchain hydrogen bonding, and the presence of ionic side groups that significantly enhance intermolecular interactions and increase thermal resistance.

Inspired by these prior results and feature considerations, we further consider that, in addition to polymer chemical structure, processing conditions may also influence Tg for polymer systems. To explore how polymer processing conditions may influence Tg and to further validate the predictive capability of the model developed in our prior work, a newly compiled dataset was used to construct an independent model, as shown in Figure 2(a). That is, the same feature set (Table 1) was used but for a new set of polymer chemistries. For comparison, the model established using the same model but using both the polymers from the previous study and this work is shown in Figure 2(b), in which only topological features were applied. In addition, as shown in Figure 2(c), the two datasets were combined to build a model incorporating as much available processing condition information as possible, so that model performance can be further assessed when processing-related factors are included.

From the results, a couple key conclusions are drawn. First, the decrease in testing accuracy with the addition of further polymers (RMSE from 33.14ºC to 46.64ºC) demonstrates the importance of data selection and that the feature selection is not global as the approach performs poorer with this new data added. This likelihood of new underlying physics added which is not captured by the feature set is further elucidated by the much lower performance (RMSE of 121.79ºC) when using only the additional polymer chemistries. Therefore, further assessment of the features could contribute to identifying the features which are missing from the additional data and to better understand the driving characteristics of the added chemistries which were missing in the prior work.

Second, the inclusion of processing parameters does not improve the model (RMSE of 48.69ºC when processing is included, as opposed to 46.64ºC) when it is not included. As mentioned, the available processing parameters were sparse, adding uncertainty to the model, but still the accuracy did not see any meaningful improvement. This largely implies that the Tg model can be considered based on chemistry, with processing having very minor impact on Tg for the majority of polymer chemistries. To further understand how the inclusion of processing information influences the model behavior, the feature importance results obtained from the two models (Figure 2b and 2c) are compared, as shown in Figure 3. In these two cases, the same set of polymers and the same gradient boosting regression model were used. The only difference is that in the second model, processing-related descriptors were incorporated in addition to the chemical descriptors. By comparing the feature importance distributions, the changes in feature weights after introducing processing conditions can be directly evaluated.

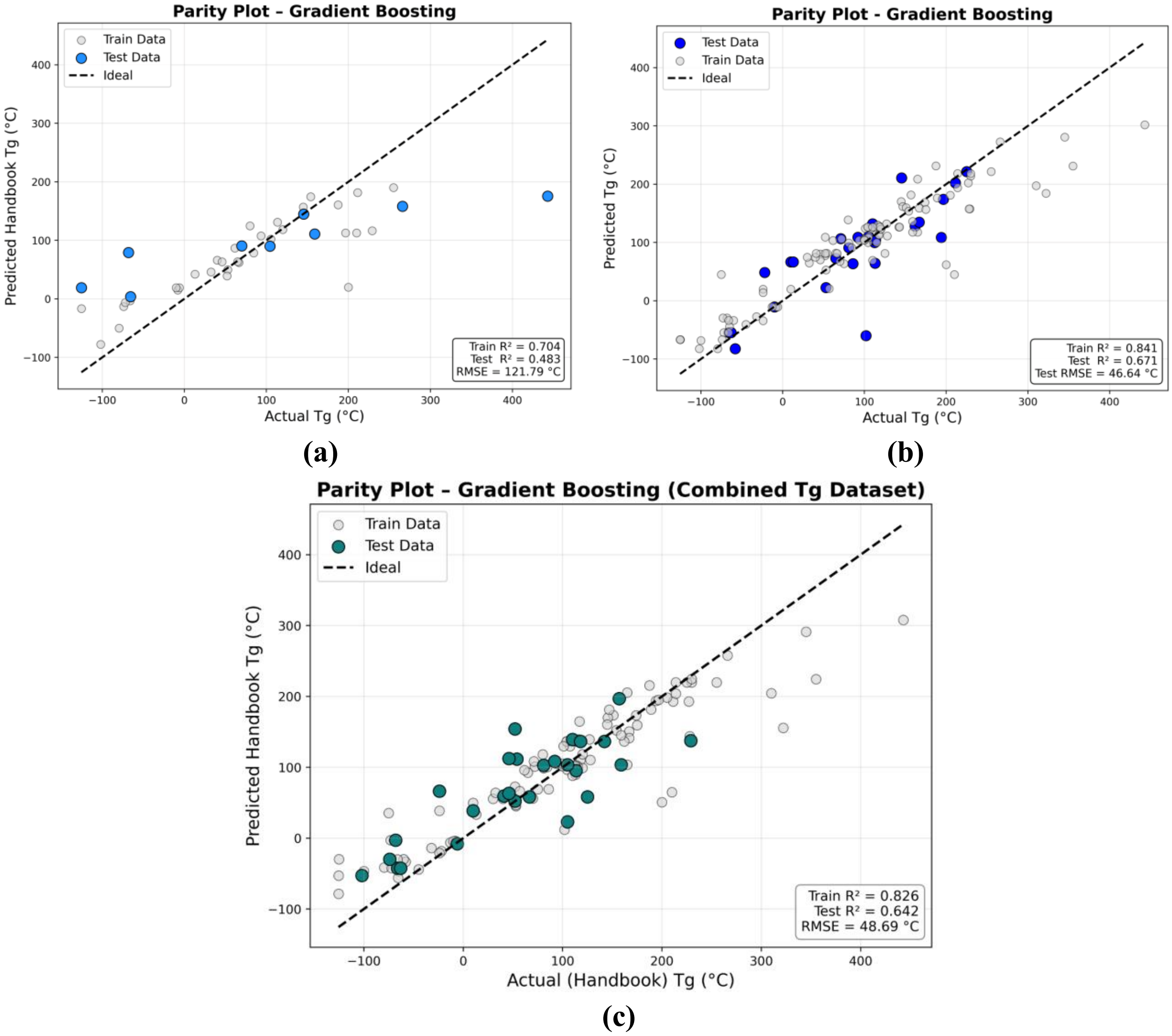

**Figure 2.** Parity plots of (a) additional polymers with available processing data, (b) the model built for all the polymers included in the prior work and this work with only topological features, and (c) the model built with the same dataset as (b) with as much as processing condition information incorporated. In all cases, gradient boosting model were used, with only minor differences in hyperparameter settings.

Although the same polymer dataset and model architecture were used, the relative importance of several descriptors changed after adding processing-related variables. This behavior is expected since gradient boosting assigns feature importance based on their relative contributions to reducing prediction error. When additional descriptors are introduced, the contribution of each feature is redistributed accordingly. Nevertheless, chemical descriptors such as 0X, N_rot, and N_CH2 remain the dominant contributors, indicating that Tg is still primarily governed by polymer chemical structure. At the same time, the importance of processing-related variables is lower than that of the chemical descriptors, indicating that processing factors play a less significant role in determining Tg. Nevertheless, their non-zero contributions suggest that they still provide complementary information and can partially account for variations in Tg that are not fully captured by chemical features alone. Overall, these results support the conclusion that Tg is mainly chemistry driven, and processing conditions act as a secondary factor rather than a primary determining factor.

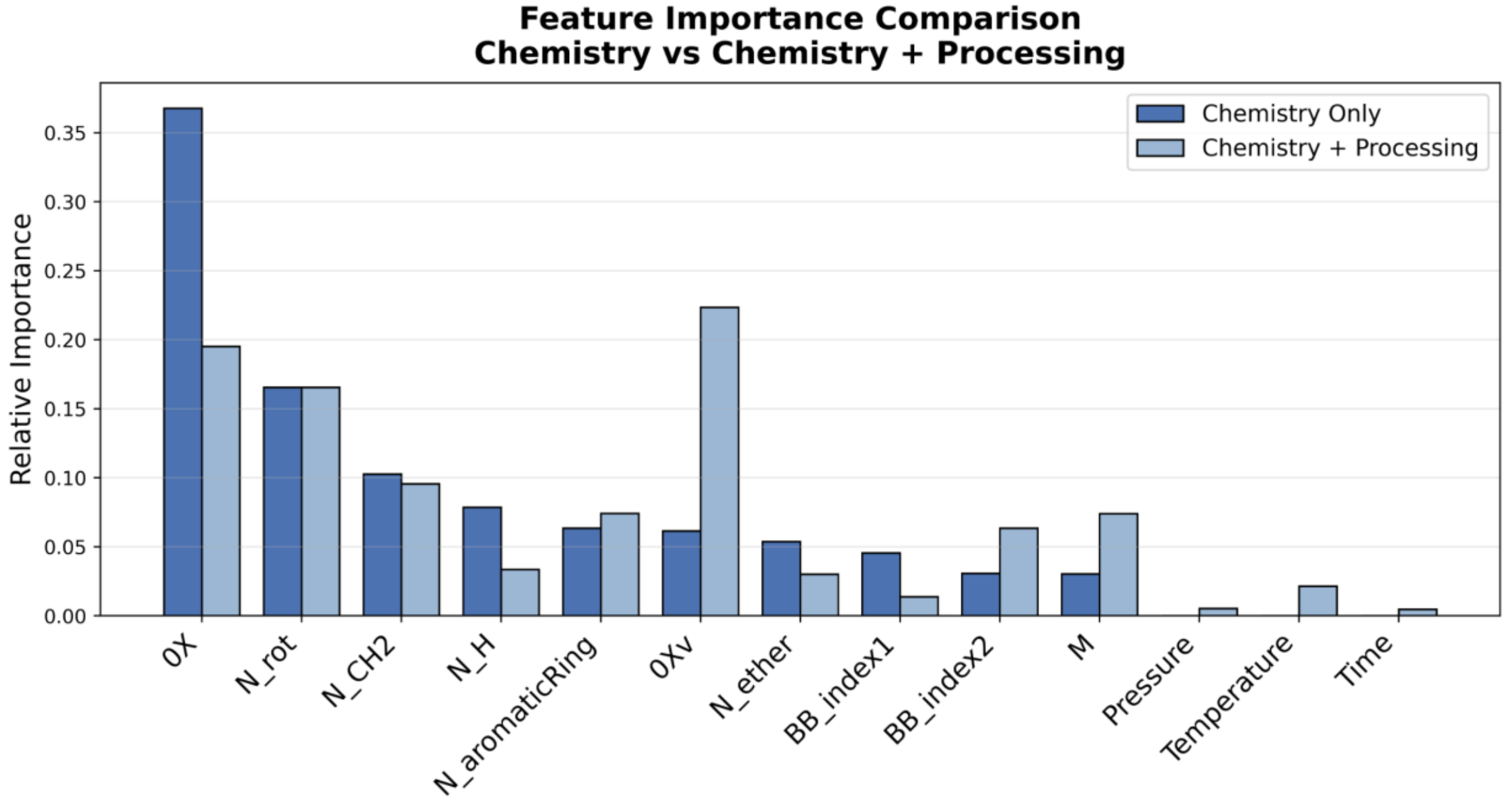

**Figure 3.** Feature importance comparison between topological features in the model and with additional processing conditions added into the dataset. The comparisons are largely similar, but with the inclusion of processing adding more importance to $^{o}X^{v}$ instead of $^{0}X$. $^{o}X^{v}$ captures valence electron effects as opposed to $^{o}X$; however, this difference is likely a statistical artifact instead of representing different underlying physics.

## IV. Discussion

The established Tg prediction model from the prior publication [1] was applied to the newly compiled polymer dataset. It is interesting to note that the overall accuracy of the model is improved over that of Figure 2, which was trained using the data that was independently calculated for Figure 4. This finding highlights two points: the generalizability of the prior model and the corresponding importance of training data, and also that processing has a minor role in determining Tg.

The model still works to screen out those chemistries which are processing dependent. Two polymers appear as outliers, for which the model predicted Tg values are noticeably lower than the reported experimental values. For polybenzimidazole, the predicted Tg is 199.72 °C, whereas the experimentally measured Tg is 443 °C. Similarly, the Tg of poly(phenylene oxide) was predicted to be 1.26 °C, while its experimental Tg is reported as 210 °C. Following this observation, we further examined possible factors that may contribute to this underprediction. Specifically, this screens out the polymers which have significant impact on Tg from processing changes versus those which are just chemistry and structure driven.

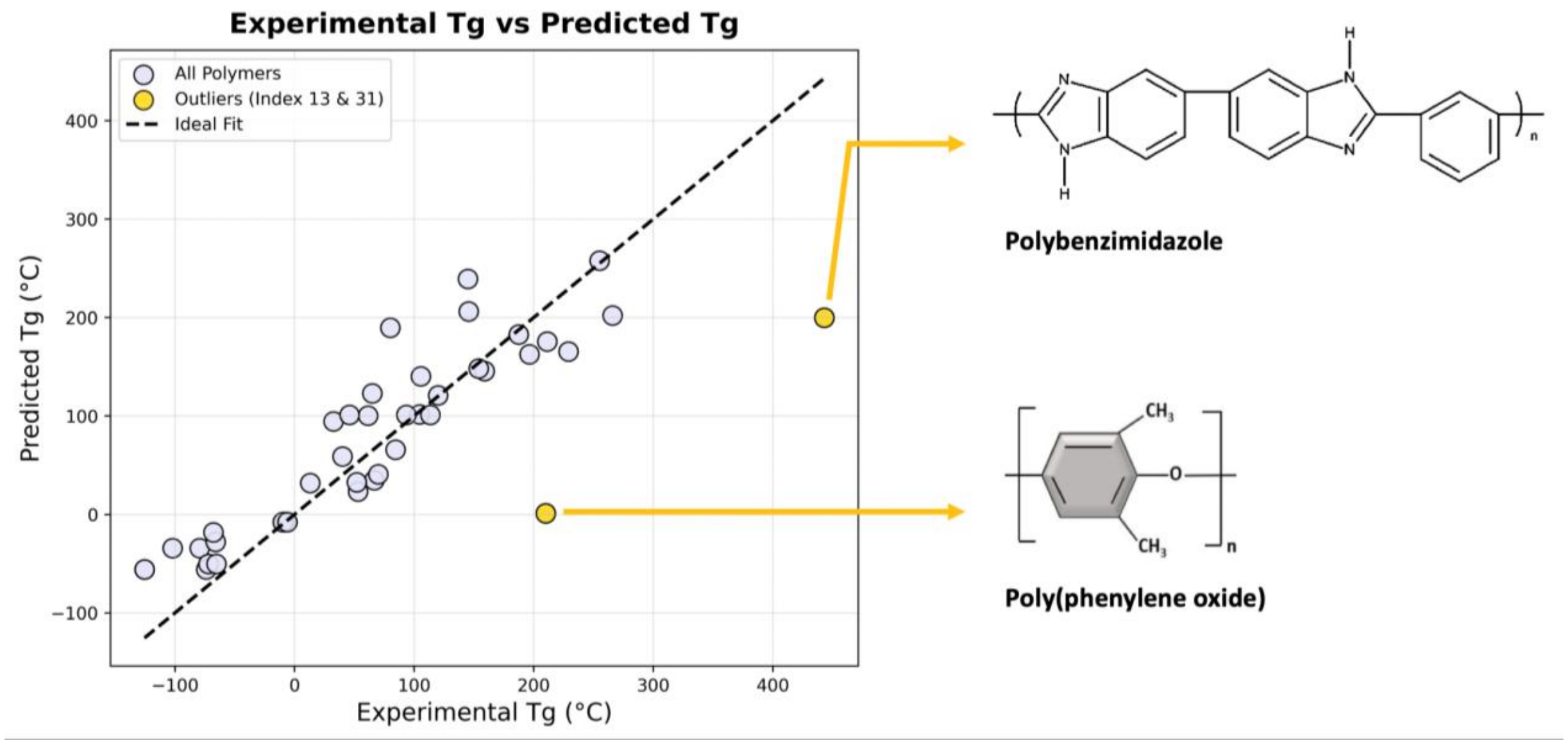

**Figure 4.** Parity plot comparing Tg using the previously established model with experimentally reported Tg values. This additional dataset was evaluated to further assess the model performance. Two outliers, polybenzimidazole and poly(phenylene oxide) are highlighted. As processing information was not included in the model, these two chemistries are identified as having Tg effected by processing, whereas the others are processing independent, in terms of Tg.

From a chemical structure perspective, polybenzimidazole has a backbone composed of fused benzimidazole rings, leading to strong aromatic interactions and extensive hydrogen bonding. The presence of a highly interconnected hydrogen-bond network, together with backbone planarity and potential π–π stacking, may not be fully captured by the topological descriptors employed in the model. Similarly, for poly(phenylene oxide), the contributions from π–π stacking interactions of the aromatic backbone may not be adequately represented, which could contribute to the observed underprediction of Tg.

From the processing condition side, polybenzimidazole cannot be melt-processed, it must be solution-cast in polyphosphoric acid (PPA), and it requires very high-temperature annealing (300–400 °C) to densify the polymer; poly(phenylene oxide) is also usually solution-cast, then annealed near Tg (about 190–200 °C) to remove residual solvent, it will produce high-density, high-rigidity films compared to melt-processed or blended poly(phenylene oxide). Importantly, as most polymers in the data were melt-processed, solution-casting process for both polybenzimidazole and poly(phenylene oxide) indicate the possibilities that processing conditions can largely affect the Tg of certain polymers. Solution casting is helpful for very dense, low–free-volume films, forming higher chain packing, fewer voids and defects, and sometimes higher Tg than melt-processed samples.

From a processing standpoint, polybenzimidazole cannot be melt-processed and is typically fabricated via solution casting in polyphosphoric acid (PPA), followed by high temperature annealing (approximately 300–400 °C) to densify the polymer structure. Poly(phenylene oxide) is also commonly prepared by solution casting and subsequently annealed near its Tg (around 190–200 °C), resulting in films with higher density and rigidity compared to melt-processed or blended polymers. In contrast, most polymers included in the handbook dataset are processed via melt-based methods. The use of solution casting for both polybenzimidazole and poly(phenylene oxide) suggests that processing conditions may play an important role in determining Tg for certain polymer systems. Solution casting can result in dense, low free volume morphologies with enhanced chain packing and reduced void content, which may lead to higher Tg values compared to melt-processed samples.

## V. Conclusion

We applied our previously established approach and topological feature development to an independent polymer dataset which includes experimentally reported Tg values, in order to quantitatively assess the role and impact of processing on Tg. The predicted Tg values using only topological features show good agreement with experimental data for most polymers, indicating stable model performance on external data and demonstrates the limited role of processing on Tg. However, we are also able to screen out those polymer chemistries which have a significant processing-Tg relationship. Two clear outliers were observed, and assessment of these identified that processing via solution casting has a significant effect on Tg. Although solution casting does not intrinsically increase Tg, it can alter polymer morphology and intermolecular interactions, leading to higher experimentally observed Tg values than those predicted based solely on chemical structure. These results indicate that while Tg is largely governed by polymer chemistry, processing conditions can play a non-negligible role for certain polymer systems, highlighting the importance of considering processing effects in future modeling efforts.